\documentclass[prl,twocolumn,twoside,showpacs,floatfix,superscriptaddress]{revtex4}

\usepackage{graphicx,color,rotating}
\usepackage{amsmath,amssymb}
\usepackage{dcolumn}
\usepackage{times,txfonts}
\usepackage{pifont}

\usepackage{dsfont}

\usepackage{pstricks}
\usepackage{pst-text}
\usepackage{pst-coil}
\usepackage{pst-node}
\usepackage{pstricks-add}

\newcommand{\symboldiamond}[1][black]{{\color{#1}\hspace{-1pt}\footnotesize\begin{turn}{45} $\blacksquare$ \end{turn}}}
\newcommand{\symboltriangle}[1][black]{{\color{#1}$\blacktriangle$}}

\newcommand{\symbolcircle}[1][black]{{\color{#1}\large$\bullet$}}

\definecolor{FGViolet}{rgb}{0.61,0.32,0.61}
\definecolor{FGDarkBlue}{rgb}{0,0,0.6}
\definecolor{FGBlue}{rgb}{0,0,0.8}
\definecolor{FGLightBlue}{rgb}{0.2, 0.6, 0.8}
\definecolor{FGGreen}{rgb}{0.2,0.7,0.2}
\definecolor{FGLightGreen}{rgb}{0.4,1,0.4}
\definecolor{FGYellow}{rgb}{1,0.95,0}
\definecolor{FGOrange}{rgb}{0.95,0.5,0.1}
\definecolor{FGRed}{rgb}{0.8,0,0}
\definecolor{FGWhite}{rgb}{1,1,1}
\definecolor{FGLightGray}{rgb}{0.8,0.8,0.8}
\definecolor{FGGray}{rgb}{0.5,0.5,0.5}
\definecolor{FGDarkGray}{rgb}{0.3,0.3,0.3}
\definecolor{FGBlack}{rgb}{0,0,0}

\newcommand{\elem}[2]{\ensuremath{{}^{#2}\text{#1}}}
\newcommand{\emax}{\ensuremath{e_{\text{max}}}}

\newcommand\CouplingTop[3]{
	\ncbar[linestyle=solid,
		   linewidth=0.5pt,
	            arm=6pt,
		   angle=90,
		   nodesep=2pt,
		   arrows=->,
		   arrowsize=2.5pt
		  ]{->}{#1}{#2}
        \naput[labelsep=3pt,
	  	  npos=1.5]
		    {{}_{#3}}
       \parbox[b][0.8cm]{0cm}{} 
}
\newcommand\CouplingBot[3]{
	\ncbar[linestyle=solid,
		   linewidth=0.5pt,
	            arm=6pt,
		   angle=-90,
		   nodesep=2pt,
		   arrows=->,
		   arrowsize=2.5pt
		  ]{->}{#1}{#2}
        \nbput[labelsep=3pt,
	  	  npos=1.5]
		     {{}_{#3}}
       \parbox[t][0.8cm]{0cm}{} 
}

\newcommand\smallSixj[6]{\left\{ \begin{smallmatrix} #1 & #2 & #3 \\ #4 & #5 & #6 \end{smallmatrix} \right\}}

\newcommand\DIAGRAMdotstyle{*}
\newcommand\DIAGRAMdotsize{15pt}
\newcommand\DIAGRAMlinewidth{2.5pt}
\newcommand\DIAGRAMarcangle{70}
\newcommand\DIAGRAMarrowsize{15pt}
\newcommand\DIAGRAM{
	\scalebox{0.2}[0.2]
	{
		\rput[rb](1,2)    { \pnode{DIAGA}  }
		\rput[rb](2,-2)    { \pnode{DIAGB}  }
		\rput[rb](2.5,0) { \psdot[dotstyle=\DIAGRAMdotstyle, dotsize=\DIAGRAMdotsize] \pnode{DIAGC}  }
		\rput[rb](3,2)    { \pnode{DIAGD}  }
		\rput[rb](4,2)    { \pnode{DIAGE}  }
		\rput[rb](5,-2)    { \pnode{DIAGF}  }
		\rput[rb](6,0)    { \psdot[dotstyle=\DIAGRAMdotstyle, dotsize=\DIAGRAMdotsize] \pnode{DIAGG}  }
		\rput[rb](7,-2)    { \pnode{DIAGH}  }
		\rput[rb](8,2)    { \pnode{DIAGI}  }
		\rput[rb](10,0)  { \psdot[dotstyle=\DIAGRAMdotstyle, dotsize=\DIAGRAMdotsize] \pnode{DIAGJ}  }
		\rput[rb](10,-2)  { \pnode{DIAGK}  }
		\ncline[linestyle=solid, linewidth=\DIAGRAMlinewidth,ArrowInside=->,ArrowInsidePos=0.5,arrowsize=\DIAGRAMarrowsize]{DIAGA}{DIAGB}
		\ncline[linestyle=solid, linewidth=\DIAGRAMlinewidth]{DIAGB}{DIAGD}
		\ncline[linestyle=solid, linewidth=\DIAGRAMlinewidth,ArrowInside=->,ArrowInsidePos=0.5,arrowsize=\DIAGRAMarrowsize]{DIAGE}{DIAGF}
		\ncline[linestyle=solid, linewidth=\DIAGRAMlinewidth]{DIAGF}{DIAGG}
		\ncline[linestyle=solid, linewidth=\DIAGRAMlinewidth]{DIAGG}{DIAGH}
		\ncline[linestyle=solid, linewidth=\DIAGRAMlinewidth]{DIAGH}{DIAGI}
		\ncline[linestyle=dashed, linewidth=\DIAGRAMlinewidth]{DIAGC}{DIAGJ}
		\psline[linestyle=solid,linewidth=\DIAGRAMlinewidth](1.5,-2)(2.5,-2)
		\psline[linestyle=solid,linewidth=\DIAGRAMlinewidth](4.5,-2)(5.5,-2)
		\psline[linestyle=solid,linewidth=\DIAGRAMlinewidth](7,-2)(10,-2)
		\ncarc[linestyle=solid,linewidth=\DIAGRAMlinewidth,arcangle=\DIAGRAMarcangle]{DIAGJ}{DIAGK}
		\ncarc[linestyle=solid,linewidth=\DIAGRAMlinewidth,arcangle=\DIAGRAMarcangle]{DIAGK}{DIAGJ}
		\parbox[b][2.5cm]{0cm}{} 
		\parbox[t][2.5cm]{0cm}{} 
	}
}
\newcommand\DIAGRAMCOUPLED{
	\scalebox{0.2}[0.2]
	{
		\rput[rb](1,2)    { \pnode{DIAGA}  }
		\rput[rb](2,-2)    { \pnode{DIAGB}  }
		\rput[rb](2.5,0) { \psdot[dotstyle=\DIAGRAMdotstyle, dotsize=\DIAGRAMdotsize] \pnode{DIAGC}  }
		\rput[rb](3,2)    { \pnode{DIAGD}  }
		\rput[rb](4,2)    { \pnode{DIAGE}  }
		\rput[rb](5,-2)    { \pnode{DIAGF}  }
		\rput[rb](6,0)    { \psdot[dotstyle=\DIAGRAMdotstyle, dotsize=\DIAGRAMdotsize] \pnode{DIAGG}  }
		\rput[rb](7,-2)    { \pnode{DIAGH}  }
		\rput[rb](8,2)    { \pnode{DIAGI}  }
		\rput[rb](10,0)  { \psdot[dotstyle=\DIAGRAMdotstyle, dotsize=\DIAGRAMdotsize] \pnode{DIAGJ}  }
		\rput[rb](10,-2)  { \pnode{DIAGK}  }
		\ncline[linestyle=solid, linewidth=\DIAGRAMlinewidth,ArrowInside=->,ArrowInsidePos=0.5,arrowsize=\DIAGRAMarrowsize]{DIAGA}{DIAGB}
		\ncline[linestyle=solid, linewidth=\DIAGRAMlinewidth]{DIAGB}{DIAGD}
		\ncline[linestyle=solid, linewidth=\DIAGRAMlinewidth,ArrowInside=->,ArrowInsidePos=0.5,arrowsize=\DIAGRAMarrowsize]{DIAGE}{DIAGF}
		\ncline[linestyle=solid, linewidth=\DIAGRAMlinewidth]{DIAGF}{DIAGG}
		\ncline[linestyle=solid, linewidth=\DIAGRAMlinewidth]{DIAGG}{DIAGH}
		\ncline[linestyle=solid, linewidth=\DIAGRAMlinewidth]{DIAGH}{DIAGI}
		\ncline[linestyle=dashed, linewidth=\DIAGRAMlinewidth]{DIAGC}{DIAGJ}
		\psline[linestyle=solid,linewidth=\DIAGRAMlinewidth](1.5,-2)(2.5,-2)
		\psline[linestyle=solid,linewidth=\DIAGRAMlinewidth](4.5,-2)(5.5,-2)
		\psline[linestyle=solid,linewidth=\DIAGRAMlinewidth](7,-2)(10,-2)
		\ncarc[linestyle=solid,linewidth=\DIAGRAMlinewidth,arcangle=\DIAGRAMarcangle]{DIAGJ}{DIAGK}
		\ncarc[linestyle=solid,linewidth=\DIAGRAMlinewidth,arcangle=\DIAGRAMarcangle]{DIAGK}{DIAGJ}
		\parbox[b][4.0cm]{0cm}{} 
		\parbox[t][2.5cm]{0cm}{} 
	}
	\ncbar[linestyle=solid, linewidth=0.5pt, arm=6pt, angle=90, nodesep=6pt, arrows=->, arrowsize=3pt ]{->}{DIAGA}{DIAGE} \naput[labelsep=3pt,npos=1.2]{{}_{JM}}
	\ncbar[linestyle=solid, linewidth=0.5pt, arm=9pt, angle=90, nodesep=6pt, arrows=->, arrowsize=3pt ]{->}{DIAGD}{DIAGI} \naput[labelsep=3pt,npos=1.5]{{}_{JM}}
}

\begin{document}

\title{\emph{Ab Initio} Calculations of Medium-Mass Nuclei with Explicit Chiral 3N Interactions}

\author{Sven Binder}
\email{sven.binder@physik.tu-darmstadt.de}
\author{Joachim Langhammer}
\author{Angelo Calci}
\affiliation{Institut f\"ur Kernphysik, Technische Universit\"at Darmstadt, 64289 Darmstadt, Germany}

\author{Petr Navr\'atil}
\affiliation{TRIUMF, 4004 Wesbrook Mall, Vancouver, British Columbia, V6T 2A3 Canada}

\author{Robert Roth}
\affiliation{Institut f\"ur Kernphysik, Technische Universit\"at Darmstadt, 64289 Darmstadt, Germany}

\date{\today}

\begin{abstract}  
We present the first \emph{ab initio} coupled-cluster calculations of medium-mass nuclei with explicit chiral three-nucleon (3N) interactions. Using a spherical formulation of coupled cluster with singles and doubles excitations including explicit 3N contributions, we study ground states of \elem{O}{16,24}, \elem{Ca}{40,48} and \elem{Ni}{56}. We employ chiral NN plus 3N interactions softened through a similarity renormalization group (SRG) transformation at the three-body level. We investigate the impact of all truncations and quantify the resulting uncertainties---this includes the contributions from triples excitations, the truncation of the set of three-body matrix elements, and the omission of SRG-induced four-body interactions. Furthermore, we assess the quality of a normal-ordering approximation of the 3N interaction beyond light nuclei. Our study points towards the predictive power of chiral Hamiltonians in the medium-mass regime.
\end{abstract}


\maketitle


The \emph{ab initio} description of medium-mass nuclei is one of the most dynamic frontiers in nuclear structure theory today---bridging the gap between accurate \emph{ab initio} calculations for light nuclei and the realm of approximate or phenomenological approaches for heavy nuclei and nuclear matter. A number of many-body methods are being developed and extended towards the medium-mass regime. Coupled-cluster (CC) theory has a pioneering role in this domain \cite{WlDe05,HaPa08,HaHj12} and other methods, like self-consistent Green's function approaches \cite{SoBa12} or the in-medium similarity renormalization group \cite{TsBo11,Herg12} are following. Extensions of the no-core shell model (NCSM) \cite{NaQu09}, like the importance-truncated NCSM \cite{Roth09,RoNa07}, are connecting the domains of light and medium-mass nuclei.

A critical ingredient for all \emph{ab initio} many-body approaches is the Hamiltonian. At present, chiral effective field theory (EFT) provides the most systematic approach to QCD-based Hamiltonians for accurate nuclear structure calculations \cite{MaEn11,EpHa09}. Already the present generation of chiral Hamiltonians, consisting of nucleon-nucleon (NN) interactions at next-to-next-to-next-to-leading order (N$^3$LO) \cite{EnMa03,EpGl05} and three-nucleon (3N) interactions at next-to-next-to-leading order (N$^2$LO) \cite{Navr07} give a very good description of p-shell nuclei as demonstrated in \emph{ab initio} NCSM calculations \cite{NaGu07,MaVa11,MaVa12,RoLa11}. Ongoing developments in this sector, e.g., regarding chiral 3N interactions at N$^3$LO \cite{BeEp08,BeEp11,SkGo11} or a $\Delta$-full formulation of chiral EFT \cite{KrEp07}, will soon provide next-generation chiral Hamiltonians with consistent NN and 3N interactions. 

The inclusion of the 3N interaction is vital to realize the predictive potential of chiral Hamiltonians, but poses a number of computational challenges. So far, there are no calculations for medium-mass nuclei that include explicit chiral 3N interactions, without resorting to approximate or even schematic reductions to effective two-body interactions \cite{HaHj12a,HaHj12}. A systematic though approximate inclusion of chiral 3N interactions for medium-mass nuclei is the normal-ordering scheme applied in Ref.~\cite{RoBi12}. 

In this Communication we present, for the first time, \emph{ab initio} coupled-cluster calculations for medium-mass nuclei including explicit 3N interactions. We have developed a spherical implementation of coupled cluster with singles and doubles excitations for three-body Hamiltonians (CCSD3B) which enables us to perform converged ground-state calculations for closed-shell nuclei with full three-body interactions. In this framework we study the ground-state energies of \elem{O}{16,24}, \elem{Ca}{40,48} and \elem{Ni}{56} using chiral NN+3N Hamiltonians softened through similarity renormalization group (SRG) transformations \cite{JuNa09,RoLa11}. We systematically address all truncations introduced in the many-body framework and the Hamiltonian and quantify the resulting uncertainties in the ground-state energies. We demonstrate that the overall uncertainty for the prediction of ground-state energies in the medium-mass regime is of the order of a few percent. Within these uncertainties the chiral NN+3N Hamiltonians used in this work predict ground-state energies that are in agreement with experiment. This is a remarkable result, since no information beyond the few-body domain ($A\leq4$) was used to fix the parameters of the chiral interactions.

\paragraph{Coupled-Cluster Method.}
In single-reference CCSD the ground state $|\Psi\rangle$ of a many-body Hamiltonian is parametrized by the exponential ansatz
\begin{eqnarray}
\label{eq:cc_ansatz}
| \Psi \rangle = e^{T} \, | \Phi \rangle
\end{eqnarray}
with $T=T_1+T_2$, where $T_n$ are excitation operators of the form
\begin{eqnarray}
T_n = 
\frac{1}{(n!)^2} 
\sum_{ a_1\dots a_n \atop i_1 \dots i_n} 
\langle a_1 \dots a_n | t_n  | i_1 \dots i_n \rangle \
a^\dagger_{ a_1} \dots a^\dagger_{ a_n} a_{ i_n} \dots a_{ i_1} 
\ ,
\end{eqnarray}
acting on a single Slater-determinant reference state $| \Phi \rangle$.

Coupled-cluster theory is conveniently formulated in terms of Hamiltonians that are normal ordered with respect to $| \Phi \rangle$. If the original Hamiltonian
\begin{eqnarray}
H= h_1 + h_2 + h_3
\end{eqnarray}
consisting of one-, two- and three-body contributions is cast into normal-ordered form,
\begin{eqnarray}\label{eq:NormalOrderedHamiltonian}
H = \langle \Phi |H | \Phi \rangle +  h^{\mathrm{NO}}_1 +h^{\mathrm{NO}}_2 + h^{\mathrm{NO}}_3 \ ,
\end{eqnarray}
contributions from the $n$-body part $h_n$ of the original Hamiltonian enter the matrix elements of the normal-ordered operators $h^{\mathrm{NO}}_k$ of particle ranks $k\le n$.

Neglecting the three-body part $h^{\mathrm{NO}}_3$ of the normal-ordered Hamiltonian (\ref{eq:NormalOrderedHamiltonian}) and thus only including contributions of the three-body interaction that have been demoted to lower particle ranks through normal-ordering leads to the NO2B approximation discussed in \cite{RoBi12,HaPa07}. In this Communication we include the complete 3N interaction, i.e., we include the normal-ordered three-body part $h^{\mathrm{NO}}_3$ beyond the NO2B approximation explicitly. The CCSD energy and amplitude equations including the full 3N interaction can be written as
\begin{eqnarray}
\label{eq:CCSDEqE}
\Delta E  &=& \Delta E_{\mathrm{NO2B}} + \langle \Phi | \,  h^{\mathrm{NO}}_3 \, e^{T}  \, | \Phi \rangle_C \\
\label{eq:CCSDEqT}
0              &=& T_{1, \mathrm{NO2B}} + \langle  \Phi^a_i |  \, h^{\mathrm{NO}}_3 \, e^{T} \, | \Phi \rangle_C \\
\label{eq:CCSDEqTT}
0               &=& T_{2, \mathrm{NO2B}} + \langle  \Phi^{ab}_{ij} | \, h^{\mathrm{NO}}_3 \, e^{T} \,   | \Phi \rangle_C 
\end{eqnarray}
where $ \Delta E_{\mathrm{NO2B}} $ and $ T_{n, \mathrm{NO2B}}$ denote the standard CCSD equations for two-body Hamiltonians \cite{ShBa09} corresponding to the NO2B approximation. Thus, the inclusion of the residual three-body operator $h^{\mathrm{NO}}_3$ generates a set of additional terms in the CCSD equations.

Expressions for (\ref{eq:CCSDEqE})-(\ref{eq:CCSDEqTT}) in $m$-scheme are presented in a factorized form in \cite{HaPa07}. We rederived these equations in a straightforward unfactorized way, resulting in more but simpler terms. 
Already for two-body Hamiltonians the basis sizes and particle numbers for $m$-scheme CC calculations are severely limited by the number of amplitudes and matrix elements that need to be handled. It is well known that the range of the CC method can be greatly extended by exploiting spherical symmetry and using an angular-momentum coupled formulation \cite{HaPa10}. We have developed such an efficient spherical implementation of CCSD3B. Proceeding along the lines of Refs.~\cite{HaPa10}, we couple the external lines of the diagrams, cut open internal lines and perform angular momentum recouplings in order to express diagrams in terms of angular-momentum coupled matrix elements of the operators involved. For example, one of the  computationally more involved contributions to the $T_2$ amplitude equations reads in $m$-scheme
\begin{eqnarray}
\lefteqn{
\DIAGRAM 
} 
\nonumber \\
&&
=
\tfrac{1}{4}  \,
P_{ab}  \,
P_{ij}  
\sum_{ c  d  e  k  l}  
\langle  k l a|h^{\mathrm{NO}}_3| c d e\rangle \,
\langle  e b| t_2 | k l \rangle \,
\langle  c|t_1| i\rangle \,
\langle  d|t_1| j\rangle
\end{eqnarray}
where we use the standard diagrammatic representation \cite{ShBa09} and the permutation operator $P_{pq} = 1-T_{pq}$ with $T_{pq}$ denoting index transpositions. The corresponding contribution in the spherical scheme is given by
\begin{eqnarray}
\lefteqn{
\DIAGRAMCOUPLED 
}
\nonumber \\
&&
=
- 
\tfrac{1}{4} \,
P_{ab}^{(J)} \,
P_{ij}^{(J)} \,
\left(  \hat{J}  \hat{\jmath}_i   \hat{\jmath}_j  \right)^{-1} \,
(-1)^{j_a+j_b-J} \,
\sum_{cdekl} \,
\sum_{J^\prime J^{\prime\prime}} \,
\hat{J}^{\prime} \,
\hat{J}^{\prime\prime} \,
\nonumber \\
&&
\times \
\smallSixj{J^\prime}{J^{\prime\prime}}{J}{j_a}{j_b}{j_e} \,
\langle \rnode{K}{k}\rnode{L}{l}\rnode{A}{\tilde a}|| h^{\mathrm{NO}}_3 || \rnode{C}{c}\rnode{D}{d}\rnode{E}{e}\rangle \,
\CouplingTop{K}{L}{J^{\prime}}
\CouplingTop{C}{D}{J}
\CouplingBot{A}{E}{J^{\prime\prime}}
\langle \rnode{E}{e} \rnode{B}{b} | {t_2} |  \rnode{K}{k} \rnode{L}{l} \rangle \CouplingTop{E}{B}{J^\prime M^\prime} \,
\CouplingTop{K}{L}{J^\prime M^\prime}
\langle \rnode{C}{\tilde c} | {t_1} | \rnode{I}{i} \rangle \CouplingTop{C}{I}{00} \,
\langle \rnode{D}{\tilde d} | {t_1} | \rnode{J}{j} \rangle \CouplingTop{D}{J}{00}
\label{eq:SphericalExampleA}
\end{eqnarray}
with $P_{pq}^{(J)} = 1-(-1)^{j_p+j_q-J} T_{pq}$ and $\hat x = \sqrt{2x+1}$. Here, indices $a,b,\dots$ represent $(nlj)$-orbitals rather than individual single-particle states. The coupling lines indicate standard angular-momentum coupling and the tilde denotes time-reversed orbitals.

In our spherical scheme, the three-body matrix elements of $h^{\mathrm{NO}}_3$ enter in a specific reduced and coupled form, given by 
\begin{eqnarray}
\lefteqn{
\langle \rnode{A}{a}\rnode{B}{b}\rnode{C}{\tilde c}|| h^{\mathrm{NO}}_3 || \rnode{D}{d}\rnode{E}{e}\rnode{F}{f}\rangle
\CouplingTop{A}{B}{J_{ab}}
\CouplingTop{D}{E}{J_{de}}
\CouplingBot{C}{F}{J_{cf}}
=
\sum_{J} (-1)^{J-j_c-J_{ab}}\hat{J}_{cf}\hat{J}^2\smallSixj{J_{ab}}{J_{de}}{J_{cf}}{j_f}{j_c}{J}
}
\nonumber \\
&& \ \ \ \ \ \times \ 
\langle[(ab)J_{ab}c]JM|h^{\mathrm{NO}}_3|[(de)J_{de}f]JM\rangle
\ ,
\end{eqnarray}
where the 3N matrix element on the r.h.s is $M$-independent. We use this specific coupling to store the three-body matrix elements in fast memory, making use of a variant of the efficient storage and retrieval schemes we have developed for $JT$-coupled three-body matrix elements \cite{RoLa11,RoCa13}. This is critical for the over-all performance of the CCSD3B calculations. In order to accelerate the convergence of the iterative solution of the CCSD3B amplitude equations, we initialize the amplitudes with the solution of the corresponding CCSD calculation in NO2B approximation. In addition to the CCSD and CCSD3B calculations we will perform $\Lambda$-CCSD(T) calculations \cite{TaRo08,TaRo08-2,HaPa10} using the NO2B Hamiltonian to assess the influence of triples excitations.

\paragraph{Hamiltonian and Basis.}

The starting point for our investigation of medium-mass nuclei are SRG-evolved chiral NN+3N Hamiltonians. We use the chiral NN interaction at N$^3$LO \cite{EnMa03} and a local form of the chiral 3N interaction at N$^2$LO \cite{Navr07}. Instead of a momentum cutoff of $500$ MeV used, e.g., in Ref. \cite{RoLa11}, we reduce the cutoff of the initial 3N interaction to $400$ MeV and choose $c_E=0.098$ to reproduce the \elem{He}{4} ground-state energy, keeping $c_D=-0.2$ \cite{RoBi12}. This cutoff reduction is motivated by the observation that SRG-induced 4N interactions have a sizable impact on ground-state energies of medium-mass nuclei, which can be reduced efficiently by reducing the cutoff of the initial 3N interaction \cite{RoLa11,RoBi12,RoCa13}. We emphasize that the following results for medium-mass nuclei from \elem{O}{16} to \elem{Ni}{56} are pure predictions.

We will employ  two types of SRG-evolved Hamiltonians: The NN+3N-full Hamiltonian starts with the initial chiral NN+3N Hamiltonian and retains all terms up to the 3N level in the SRG evolution, the NN+3N-induced Hamiltonian omits the chiral 3N interaction from the initial Hamiltonian, but keeps all induced 3N terms throughout the evolution. In addition, we use a range of flow parameters $\alpha$ in order to assess the role of SRG-induced contributions beyond the three-body level \cite{RoLa11}.

The underlying single-particle basis is a harmonic-oscillator basis truncated in the principal oscillator quantum number $2n+l = e \leq e_{\max}$ and we go up to $e_{\max}=12$. We perform Hartree-Fock (HF) calculations including the 3N interaction for each set of basis parameters to obtain an optimized single-particle basis and stabilize the convergence of the CC iterations. The normal-ordering is done consistently, i.e., with respect to the HF reference state. At the moment it is not possible to include all three-body matrix elements that would appear in the larger bases, we are limited to three-body matrix elements with $e_1+e_2+e_3 \leq E_{3\max} = 14$. We will discuss the impact of this additional cut in detail later on. We stress that the exact treatment of the isospin dependence of the two- and three-body matrix elements in the HF basis is crucial. It is generated by isospin-dependence of the HF single-particle wavefunctions, although the 3N operator used here is isospin symmetric. In order to avoid a drastic increase of storage needed for the HF three-body matrix elements, we perform the transformation to the HF basis on the fly.

\paragraph{Results.}

\begin{figure}[b]
\includegraphics[width=0.85\columnwidth]{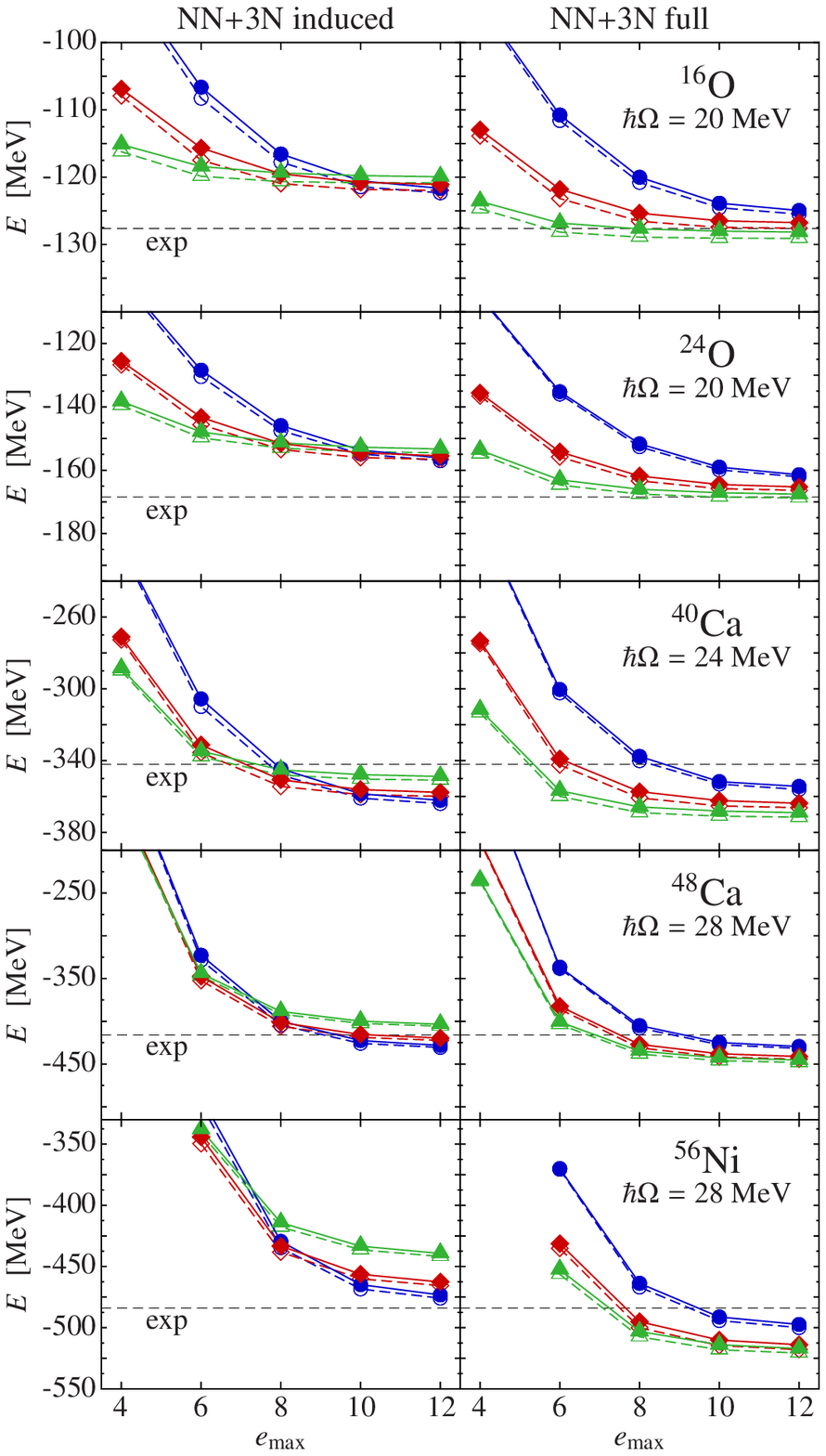}
\caption{(color online) Ground-state energies for \elem{O}{16,24}, \elem{Ca}{40,48} and \elem{Ni}{56} as function of $e_{\max}$ for the two types of Hamiltonians (see column headings)  using CCSD3B (solid lines) and the NO2B  approximation (dashed lines) for a range of flow parameters: $\alpha=0.02\,\text{fm}^4$ (\symbolcircle[FGBlue]), $0.04\,\text{fm}^4$ (\symboldiamond[FGRed]), $0.08\,\text{fm}^4$ (\symboltriangle[FGGreen]). }
\label{fig:1}
\end{figure}

We present first results of CC calculations with explicit 3N interactions for the ground-state energies of \elem{O}{16,24}, \elem{Ca}{40,48}, and \elem{Ni}{56}. Step by step, we will quantify the uncertainties resulting from various truncations in the many-body treatment. First we contrast CCSD3B calculations with CCSD using the NO2B approximation for the Hamiltonian. For both we discuss basis-space convergence in terms of $e_{\max}$ and oscillator frequency $\hbar\Omega$. We then study the influence of the three-body truncation $E_{3\max}$. Finally, we include non-iterative triples corrections at the level of $\Lambda$-CCSD(T) with the NO2B Hamiltonian. In all cases we vary the flow parameter $\alpha$ over a wide range to study the impact of induced many-body terms.

\begin{table}[t]												
\caption{Summary of CCSD and $\Lambda$-CCSD(T) ground-state energies in MeV for the NN+3N-induced Hamiltonian  for a subset of $\alpha$ values computed at optimum oscillator frequency $\hbar \Omega=20$ MeV for \elem{O}{16,24}, $\hbar \Omega=24$ MeV for \elem{Ca}{40}, and $\hbar \Omega=28$ MeV for \elem{Ca}{48}, \elem{Ni}{56}. The number in parenthesis denotes the $E_{3\max}$ cut in the 3N Hamiltonian. The last column gives $\emax$-extrapolated values, the other columns are for $e_{\max}=12$.} \label{tab:NN3Nind}												
\begin{ruledtabular}												
\begin{tabular}{ccccccc}												
NN+3N		&					& CCSD	         &  \multicolumn{2}{c}{CCSD}    &  \multicolumn{2}{c}{$\Lambda$-CCSD(T)} \\	   
induced	          &$\alpha [\text{fm}^4]$    & 3B	         &  NO2B              & NO2B              & NO2B                             &  NO2B$^{\infty}$  \\	   
			&				          & (12)                 & (12)   		  & (14) 		   & (14)                 		&  (14)  \\

\hline			
\elem{O}{16}	&	0.02	&	-121.6	&	-122.3	&	-121.7	&	-126.1	&	-126.5	\\
	&	0.04	&	-121.1	&	-121.9	&	-121.6	&	-124.4	&	-124.4	\\
	&	0.08	&	-119.9	&	-120.8	&	-120.8	&	-122.4	&	-122.4	\\
\elem{O}{24}	&	0.02	&	-156.1	&	-157.0	&	-155.8	&	-162.7	&	-163.7	\\
	&	0.04	&	-155.4	&	-156.6	&	-155.9	&	-160.2	&	-160.4	\\
	&	0.08	&	-153.3	&	-154.5	&	-154.4	&	-156.9	&	-157.1	\\
\elem{Ca}{40}	&	0.02	&	-362.1	&	-363.9	&	-360.5	&	-374.3	&	-375.5	\\
	&	0.04	&	-357.7	&	-359.9	&	-358.1	&	-366.3	&	-366.6	\\
	&	0.08	&	-348.8	&	-350.9	&	-350.8	&	-355.3	&	-355.5	\\
\elem{Ca}{48}	&	0.02	&	-428.1	&	-430.5	&	-425.3	&	-442.3	&	-443.9	\\
	&	0.04	&	-419.6	&	-422.3	&	-420.0	&	-429.6	&	-430.9	\\
	&	0.08	&	-403.4	&	-405.7	&	-406.8	&	-411.9	&	-413.3	\\
\elem{Ni}{56}	&	0.02	&	-473.2	&	-475.8	&	-464.3	&	-487.5	&	-489.8	\\
	&	0.04	&	-462.6	&	-465.6	&	-458.0	&	-472.4	&	-474.2	\\
	&	0.08	&	-439.3	&	-441.8	&	-439.3	&	-448.4	&	-450.4	\\
\end{tabular}												
\end{ruledtabular}												
\end{table}												
\begin{table}												
\caption{Same as in Tab. \ref{tab:NN3Nind} for the NN+3N-full Hamiltonian.}												
\label{tab:NN3Nfull}												
\begin{ruledtabular}												
\begin{tabular}{ccccccc}												
NN+3N		&					& CCSD	         &  \multicolumn{2}{c}{CCSD}    &  \multicolumn{2}{c}{$\Lambda$-CCSD(T)} \\	   
full		          &$\alpha [\text{fm}^4]$    & 3B	         &  NO2B              & NO2B              & NO2B                             &  NO2B$^{\infty}$  \\	   
			&				          & (12)                 & (12)   		  & (14) 		   & (14)                 		&  (14)  \\
\hline
\elem{O}{16}	&	0.02	&	-124.9	&	-125.4	&	-124.5	&	-129.9	&	-130.4	\\
			&	0.04	&	-126.8	&	-127.6	&	-127.1	&	-130.8	&	-130.8	\\
			&	0.08	&	-128.2	&	-129.1	&	-129.0	&	-131.2	&	-131.2	\\
\elem{O}{24}	&	0.02	&	-161.4	&	-162.0	&	-160.4	&	-168.3	&	-169.4	\\
			&	0.04	&	-165.3	&	-166.4	&	-165.5	&	-170.7	&	-170.9	\\
			&	0.08	&	-167.6	&	-168.8	&	-168.6	&	-171.6	&	-171.7	\\
\elem{Ca}{40}	&	0.02	&	-354.4	&	-356.0	&	-352.1	&	-370.6	&	-371.7	\\
			&	0.04	&	-363.8	&	-366.2	&	-364.3	&	-376.4	&	-376.7	\\
			&	0.08	&	-369.0	&	-371.5	&	-371.3	&	-378.2	&	-378.4	\\
\elem{Ca}{48}	&	0.02	&	-429.4	&	-431.6	&	-426.7	&	-449.5	&	-450.9	\\
			&	0.04	&	-441.2	&	-444.3	&	-441.9	&	-456.0	&	-457.0	\\
			&	0.08	&	-445.3	&	-448.2	&	-448.3	&	-453.5	&	-456.7	\\
\elem{Ni}{56}	&	0.02	&	-497.3	&	-499.9	&	-490.9	&	-521.7	&	-523.4	\\
			&	0.04	&	-513.9	&	-517.9	&	-511.4	&	-530.9	&	-531.8	\\
			&	0.08	&	-517.0	&	-520.7	&	-517.9	&	-528.4	&	-529.2	\\
\end{tabular}												
\end{ruledtabular}												
\end{table}					

\begin{figure}[b]
\includegraphics[width=0.9\columnwidth]{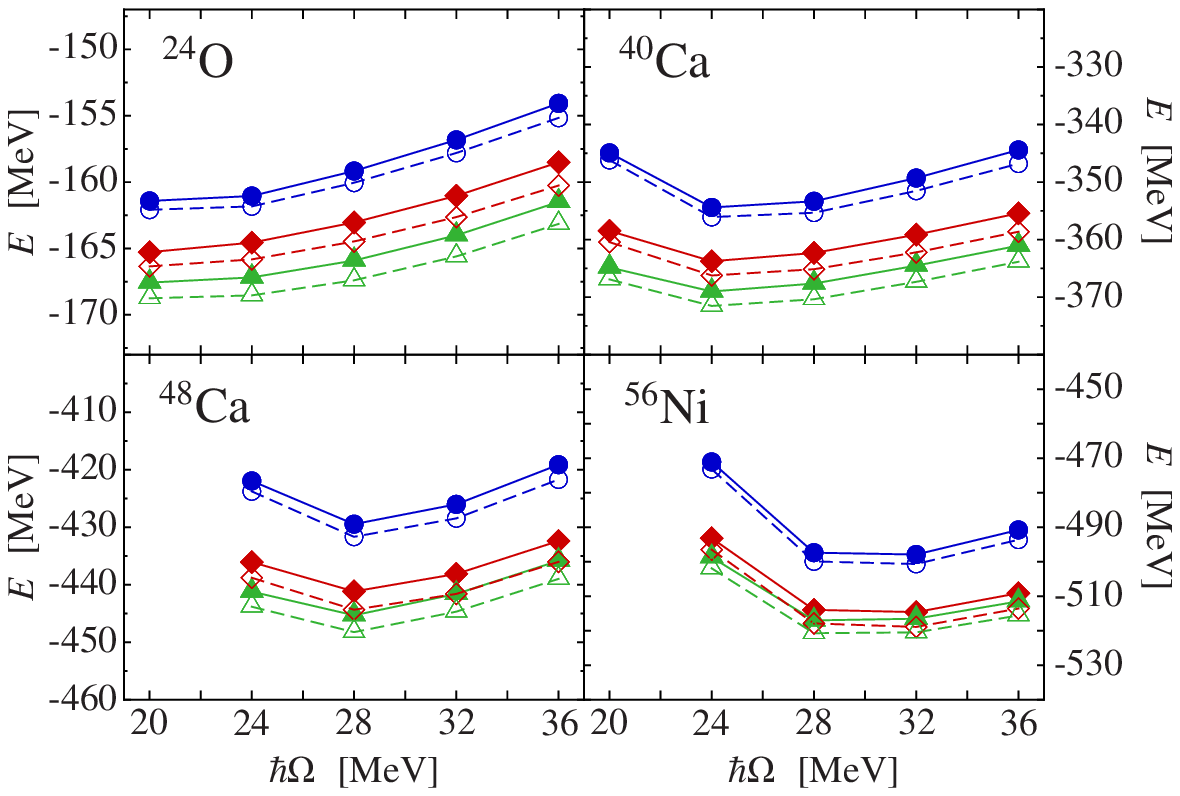}
\caption{ 
(color online) Ground-state energies for \elem{O}{24}, \elem{Ca}{40,48} and \elem{Ni}{56} as function of $\hbar\Omega$ at $e_{\max}=12$ for the NN+3N-full Hamiltonian using CCSD3B (solid lines) and the NO2B approximation  (dashed lines)  for a range of flow parameters: $\alpha=0.02\,\text{fm}^4$ (\symbolcircle[FGBlue]), $0.04\,\text{fm}^4$ (\symboldiamond[FGRed]), $0.08\,\text{fm}^4$ (\symboltriangle[FGGreen]). 
}
\label{fig:2}
\end{figure}

Figure \ref{fig:1} shows a comparison of CCSD3B calculations using the complete 3N interaction with CCSD using the NO2B approximation for the ground-state energies of \elem{O}{16,24}, \elem{Ca}{40,48}, and \elem{Ni}{56} as function of the basis size $e_{\max}$ for, both, the NN+3N-induced and the NN+3N-full Hamiltonian with three different values of the SRG flow parameter. The oscillator frequencies correspond to the energy minima in the largest basis spaces (cf. Fig. \ref{fig:2}). The numerical values of the ground-state energies for the largest basis sets are also summarized in Tabs. \ref{tab:NN3Nind} and \ref{tab:NN3Nfull}. The first observation is that we are able to converge or come sufficiently close to convergence with respect to the basis size $e_{\max}$ in practically all cases. The second observation is that the NO2B works extremely well for all cases: For \elem{O}{16} the largest deviation from the full CCSD3B results is $0.9$ MeV or $0.8\%$, for \elem{Ni}{56} it is $4$ MeV or $0.8$\% across all Hamiltonians considered here. Given that the computational cost for the CCSD3B calculations is two orders of magnitude higher than for CCSD and that the accuracy we target for many-body calculations in this mass range is not better than $1$\%, the NO2B approximation constitutes a very efficient tool.

The quality of the NO2B approximation is confirmed in Fig. \ref{fig:2}, where it is compared to CCSD3B using the NN+3N-full Hamiltonian and the largest basis set as function of the oscillator frequency. The accuracy of the NO2B approximation is largely independent of $\hbar\Omega$. Note that the effect of the residual 3N interaction beyond the NO2B approximation is always repulsive, i.e., of the same sign as the complete 3N contribution composed of induced and evolved initial 3N terms \cite{RoBi12}. 

The fact that the SRG evolution in the 3N sector is performed in a finite model space of harmonic oscillator Jacobi states \cite{RoLa11,RoCa13} leads to additional uncertainties at low frequencies $\hbar\Omega$. By varying the size of SRG model space \cite{RoCa13} we estimate the uncertainties at the optimal frequency to be much smaller than 1\% for all nuclei except \elem{Ni}{56}, where they reach the 1\% level. For smaller frequencies, however, the truncation of the SRG model space leads to more significant effects---the increase of the ground-state energies of \elem{Ca}{40} and beyond at the lowest frequencies shown in Fig. \ref{fig:2} is partly due to this. 
 
Next we address the $E_{3\max}$ cut used in the 3N matrix elements for technical reasons. In Tabs. \ref{tab:NN3Nind} and \ref{tab:NN3Nfull} the CCSD results using the NO2B approximation with $E_{3\max}=12$ and $14$ are compared. We find that the influence of this cut grows with increasing particle number and decreasing flow parameter $\alpha$. For the softest interaction with $\alpha=0.08\,\text{fm}^4$ the cut is completely irrelevant up to \elem{Ca}{40}, only for \elem{Ni}{56} we observe a $0.5$\% change in the ground-state energy. For less evolved interactions the effect increases and reaches about $1\%$ up to \elem{Ca}{48} and about $2$\% for \elem{Ni}{56}. For the description of still heavier nuclei or the use of bare 3N interactions one will have to improve on this truncation in order to reach accurate results.

\begin{figure}[b]
\includegraphics[width=0.85\columnwidth]{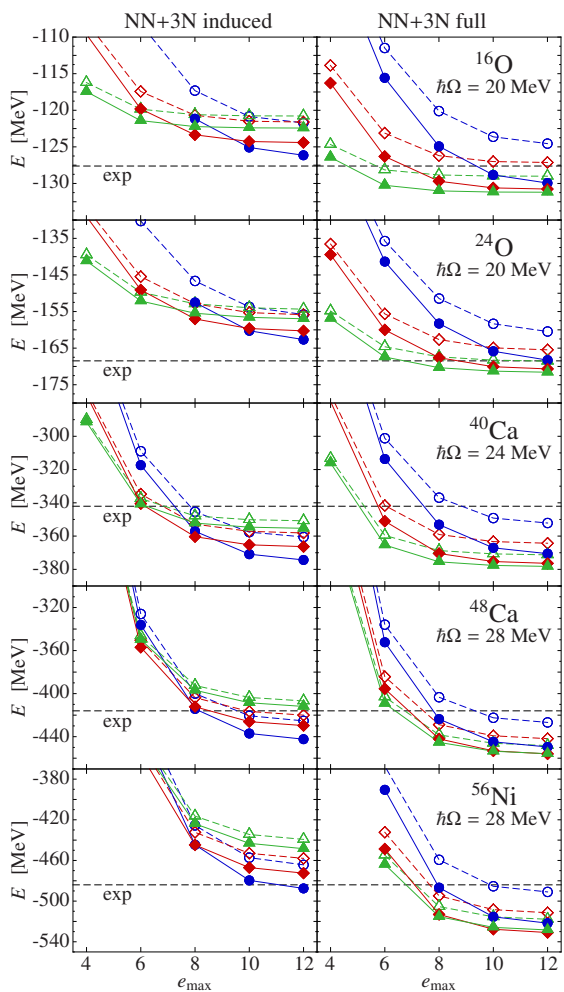}
\caption{
(color online) $\Lambda$-CCSD(T) (solid lines) and CCSD (dashed lines) ground-state energies for \elem{O}{16,24}, \elem{Ca}{40,48} and \elem{Ni}{56} as function of $e_{\max}$ for the two types of Hamiltonians (see column headings)  using the NO2B  approximation for flow parameters: $\alpha=0.02\,\text{fm}^4$ (\symbolcircle[FGBlue]), $0.04\,\text{fm}^4$ (\symboldiamond[FGRed]), $0.08\,\text{fm}^4$ (\symboltriangle[FGGreen]).
}
\label{fig:3}
\end{figure}

As the final model-space related truncation, we discuss the truncation of the excitation operators in the coupled cluster ansatz \eqref{eq:cc_ansatz}. In Fig. \ref{fig:3} we compare the results of CCSD and $\Lambda$-CCSD(T) calculations for all nuclei and Hamiltonians in NO2B approximation as function of the basis truncation $e_{\max}$ for $E_{3\max}=14$. The results for $e_{\max}=12$ are also summarized in Tabs. \ref{tab:NN3Nind} and \ref{tab:NN3Nfull}. Again we observe a systematic dependence on the flow parameter $\alpha$. For the softest interactions with $\alpha=0.08\,\text{fm}^4$ the inclusion of triples excitations lowers the ground-state energies by $1.5$ to $2\%$ for all nuclei and both Hamiltonians. For $\alpha=0.02\,\text{fm}^4$ the difference increases to about $4$ to $6\%$. If we conservatively consider the triples contribution as a measure for the inherent uncertainty due to the truncation of the cluster operator, then this is the largest uncertainty so far. 

Finally, we quantify the uncertainty due to the omission of the SRG-induced four- and more-nucleon interactions through the $\alpha$-dependence of the CCSD and $\Lambda$-CCSD(T) results shown in Fig. \ref{fig:3} and Tabs. \ref{tab:NN3Nind} and \ref{tab:NN3Nfull}. First of all, we note that missing many-body terms of the Hamiltonian are of opposite sign but of the same order of magnitude for the NN+3N-induced and the NN+3N-full Hamiltonian. For the NN+3N-induced Hamiltonian the ground-state energies for $\alpha=0.02\,\text{fm}^4$ (harder interaction) are systematically lower than for $\alpha=0.08\,\text{fm}^4$ (softer interaction). Furthermore, the energy spread over this range of flow parameters is smaller for CCSD and larger for $\Lambda$-CCSD(T). For \elem{Ca}{40}, e.g., the spread amounts to 3\% of the ground-state energy in CCSD and 5\% in $\Lambda$-CCSD(T). The pattern is reversed for the NN+3N-full Hamiltonian. The ground-state energies for $\alpha=0.02\,\text{fm}^4$ are systematically above the energies for $\alpha=0.08\,\text{fm}^4$ and the energy spread is reduced by including the triples correction. For \elem{Ca}{40} the relative energy spread is 5\% for the CCSD and $2\%$ for $\Lambda$-CCSD(T). 
Note that these $\alpha$-dependencies are distorted by the influence of the $E_{3\max}$-truncation. With increasing $E_{3\max}$ the $\Lambda$-CCSD(T) ground-state energies will move up for harder interactions and, thus, the apparent $\alpha$-dependence will be reduced for NN+3N-induced and increased for NN+3N-full. 

If we compare the ground-state energies throughout the set of nuclei discussed here with experiment (cf. Fig. \ref{fig:3}), keeping in mind the uncertainties we discussed above, then the agreement is remarkable. We stress that no information beyond $A=4$ was used to constrain the Hamiltonian, so obtaining the correct binding systematics for medium-mass nuclei is far from trivial. Though the impact of the initial chiral 3N interaction is moderate on the scales shown in Fig. \ref{fig:3}, it is important to obtain the correction binding-energy systematics along isotopic chains. In contrast, the effect of the SRG-induced 3N interactions is huge and their inclusion is mandatory --- for \elem{Ni}{56} and $\alpha=0.08\,\text{fm}^4$ a calculation with only SRG-evolved NN interactions yields an unphysical ground-state energy of about $-950$ MeV.

\paragraph{Conclusions.}

We have presented the first \emph{ab initio} calculations for nuclei in the medium-mass regime with inclusion of explicit 3N interactions. On this baseline we quantify the effects of all truncations in the many-body approach, i.e., the cluster-rank of the CC ansatz, the single-particle truncation $e_{\max}$, the truncation of the 3N matrix elements $E_{3\max}$, and the optional omission of residual normal-ordered 3N terms. For all truncations we clearly benefit from the prediagonalization of the Hamiltonian through the SRG evolution---for Hamiltonians with $\alpha=0.08\,\text{fm}^4$ the uncertainties due to the $E_{3\max}$ truncation and the NO2B approximation are at the level of $1\%$ or below and the effect of the triples correction is only $2\%$. In particular the impact of the $E_{3\max}$ truncation and the triples corrections increase rapidly when going to harder interactions. The omission of SRG-induced four- and many-body terms, i.e. the truncation of the particle rank of the Hamiltonian, introduces an uncertainty at the level of a few \% and thus limits the over-all accuracy of the approach. A reduction or elimination of this uncertainty will be a prime goal of future studies. Finally, our results point towards the predictive power of chiral NN+3N Hamiltonians for medium-mass nuclei.

\paragraph{Acknowledgments.}

Numerical calculations have been performed at the J\"ulich Supercomputing Centre, at the LOEWE-CSC Frankfurt, and at the National Energy Research Scientific Computing Center supported by the Office of Science of the U.S. Department of Energy under Contract No. DE-AC02-05CH11231.  
Supported by the Deutsche Forschungsgemeinschaft through contract SFB 634, by the Helmholtz International Center for FAIR within the LOEWE program of the State of Hesse, and the BMBF through contracts 06DA9040I and 06DA7047I. P.N. acknowledges support from Natural Sciences and Engineering Research Council of Canada (NSERC) Grant No. 401945-2011. 


\begin{thebibliography}{30}
\expandafter\ifx\csname natexlab\endcsname\relax\def\natexlab#1{#1}\fi
\expandafter\ifx\csname bibnamefont\endcsname\relax
  \def\bibnamefont#1{#1}\fi
\expandafter\ifx\csname bibfnamefont\endcsname\relax
  \def\bibfnamefont#1{#1}\fi
\expandafter\ifx\csname citenamefont\endcsname\relax
  \def\citenamefont#1{#1}\fi
\expandafter\ifx\csname url\endcsname\relax
  \def\url#1{\texttt{#1}}\fi
\expandafter\ifx\csname urlprefix\endcsname\relax\def\urlprefix{URL }\fi
\providecommand{\bibinfo}[2]{#2}
\providecommand{\eprint}[2][]{\url{#2}}

\bibitem[{\citenamefont{Wloch et~al.}(2005)\citenamefont{Woch, Dean, Gour,
  Hjorth-Jensen, Kowalski, Papenbrock, and Piecuch}}]{WlDe05}
\bibinfo{author}{\bibfnamefont{M.}~\bibnamefont{W\l{}och}},
  \bibinfo{author}{\bibfnamefont{D.}~\bibnamefont{Dean}},
  \bibinfo{author}{\bibfnamefont{J.}~\bibnamefont{Gour}},
  \bibinfo{author}{\bibfnamefont{M.}~\bibnamefont{Hjorth-Jensen}},
  \bibinfo{author}{\bibfnamefont{K.}~\bibnamefont{Kowalski}},
  \bibinfo{author}{\bibfnamefont{T.}~\bibnamefont{Papenbrock}},
  \bibnamefont{and} \bibinfo{author}{\bibfnamefont{P.}~\bibnamefont{Piecuch}},
  \bibinfo{journal}{Phys. Rev. Lett.} \textbf{\bibinfo{volume}{94}},
  \bibinfo{pages}{212501} (\bibinfo{year}{2005}).

\bibitem[{\citenamefont{Hagen et~al.}(2008)\citenamefont{Hagen, Papenbrock,
  Dean, and Hjorth-Jensen}}]{HaPa08}
\bibinfo{author}{\bibfnamefont{G.}~\bibnamefont{Hagen}},
  \bibinfo{author}{\bibfnamefont{T.}~\bibnamefont{Papenbrock}},
  \bibinfo{author}{\bibfnamefont{D.~J.} \bibnamefont{Dean}}, \bibnamefont{and}
  \bibinfo{author}{\bibfnamefont{M.}~\bibnamefont{Hjorth-Jensen}},
  \bibinfo{journal}{Phys. Rev. Lett.} \textbf{\bibinfo{volume}{101}},
  \bibinfo{pages}{092502} (\bibinfo{year}{2008}).

\bibitem[{\citenamefont{Hagen et~al.}(2012{\natexlab{a}})\citenamefont{Hagen,
  Hjorth-Jensen, Jansen, Machleidt, and Papenbrock}}]{HaHj12}
\bibinfo{author}{\bibfnamefont{G.}~\bibnamefont{Hagen}},
  \bibinfo{author}{\bibfnamefont{M.}~\bibnamefont{Hjorth-Jensen}},
  \bibinfo{author}{\bibfnamefont{G.~R.} \bibnamefont{Jansen}},
  \bibinfo{author}{\bibfnamefont{R.}~\bibnamefont{Machleidt}},
  \bibnamefont{and}
  \bibinfo{author}{\bibfnamefont{T.}~\bibnamefont{Papenbrock}},
  \bibinfo{journal}{Phys. Rev. Lett.} \textbf{\bibinfo{volume}{109}},
  \bibinfo{pages}{032502} (\bibinfo{year}{2012}{\natexlab{a}}).

\bibitem[{\citenamefont{Soma et~al.}(2012)\citenamefont{Soma, Barbieri, and
  Duguet}}]{SoBa12}
\bibinfo{author}{\bibfnamefont{V.}~\bibnamefont{Soma}},
  \bibinfo{author}{\bibfnamefont{C.}~\bibnamefont{Barbieri}}, \bibnamefont{and}
  \bibinfo{author}{\bibfnamefont{T.}~\bibnamefont{Duguet}},
  \bibinfo{journal}{arXiv:1208.2472v1}  (\bibinfo{year}{2012}).

\bibitem[{\citenamefont{Tsukiyama et~al.}(2011)\citenamefont{Tsukiyama, Bogner,
  and Schwenk}}]{TsBo11}
\bibinfo{author}{\bibfnamefont{K.}~\bibnamefont{Tsukiyama}},
  \bibinfo{author}{\bibfnamefont{S.~K.} \bibnamefont{Bogner}},
  \bibnamefont{and} \bibinfo{author}{\bibfnamefont{A.}~\bibnamefont{Schwenk}},
  \bibinfo{journal}{Phys. Rev. Lett.} \textbf{\bibinfo{volume}{106}},
  \bibinfo{pages}{222502} (\bibinfo{year}{2011}).

\bibitem[{\citenamefont{Hergert et~al.}()\citenamefont{Hergert, Bogner, Binder,
  Calci, Langhammer, and Roth}}]{Herg12}
\bibinfo{author}{\bibfnamefont{H.}~\bibnamefont{Hergert}},
  \bibinfo{author}{\bibfnamefont{S.~K.} \bibnamefont{Bogner}},
  \bibinfo{author}{\bibfnamefont{S.}~\bibnamefont{Binder}},
  \bibinfo{author}{\bibfnamefont{A.}~\bibnamefont{Calci}},
  \bibinfo{author}{\bibfnamefont{J.}~\bibnamefont{Langhammer}},
  \bibinfo{author}{\bibfnamefont{A.}~\bibnamefont{Schwenk}},
  \bibnamefont{and} \bibinfo{author}{\bibfnamefont{R.}~\bibnamefont{Roth}},
  \bibinfo{note}{in preparation}.

\bibitem[{\citenamefont{Navr\'atil et~al.}(2009)\citenamefont{Navr\'atil,
  Quaglioni, Stetcu, and Barrett}}]{NaQu09}
\bibinfo{author}{\bibfnamefont{P.}~\bibnamefont{Navr\'atil}},
  \bibinfo{author}{\bibfnamefont{S.}~\bibnamefont{Quaglioni}},
  \bibinfo{author}{\bibfnamefont{I.}~\bibnamefont{Stetcu}}, \bibnamefont{and}
  \bibinfo{author}{\bibfnamefont{B.}~\bibnamefont{Barrett}},
  \bibinfo{journal}{J. Phys. G: Nucl. Part. Phys.}
  \textbf{\bibinfo{volume}{36}}, \bibinfo{pages}{083101}
  (\bibinfo{year}{2009}).

\bibitem[{\citenamefont{Roth}(2009)}]{Roth09}
\bibinfo{author}{\bibfnamefont{R.}~\bibnamefont{Roth}}, \bibinfo{journal}{Phys.
  Rev. C} \textbf{\bibinfo{volume}{79}}, \bibinfo{pages}{064324}
  (\bibinfo{year}{2009}).

\bibitem[{\citenamefont{Roth and Navr\'atil}(2007)}]{RoNa07}
\bibinfo{author}{\bibfnamefont{R.}~\bibnamefont{Roth}} \bibnamefont{and}
  \bibinfo{author}{\bibfnamefont{P.}~\bibnamefont{Navr\'atil}},
  \bibinfo{journal}{Phys. Rev. Lett.} \textbf{\bibinfo{volume}{99}},
  \bibinfo{pages}{092501} (\bibinfo{year}{2007}).

\bibitem[{\citenamefont{Machleidt and Entem}(2011)}]{MaEn11}
\bibinfo{author}{\bibfnamefont{R.}~\bibnamefont{Machleidt}} \bibnamefont{and}
  \bibinfo{author}{\bibfnamefont{D.~R.} \bibnamefont{Entem}},
  \bibinfo{journal}{Phys. Rep.} \textbf{\bibinfo{volume}{503}},
  \bibinfo{pages}{1} (\bibinfo{year}{2011}).

\bibitem[{\citenamefont{Epelbaum et~al.}(2009)\citenamefont{Epelbaum, Hammer,
  and Mei\ss{}ner}}]{EpHa09}
\bibinfo{author}{\bibfnamefont{E.}~\bibnamefont{Epelbaum}},
  \bibinfo{author}{\bibfnamefont{H.-W.} \bibnamefont{Hammer}},
  \bibnamefont{and} \bibinfo{author}{\bibfnamefont{U.-G.}
  \bibnamefont{Mei\ss{}ner}}, \bibinfo{journal}{Rev. Mod. Phys.}
  \textbf{\bibinfo{volume}{81}}, \bibinfo{pages}{1773} (\bibinfo{year}{2009}).

\bibitem[{\citenamefont{Entem and Machleidt}(2003)}]{EnMa03}
\bibinfo{author}{\bibfnamefont{D.~R.} \bibnamefont{Entem}} \bibnamefont{and}
  \bibinfo{author}{\bibfnamefont{R.}~\bibnamefont{Machleidt}},
  \bibinfo{journal}{Phys. Rev. C} \textbf{\bibinfo{volume}{68}},
  \bibinfo{pages}{041001(R)} (\bibinfo{year}{2003}).

\bibitem[{\citenamefont{Epelbaum et~al.}(2005)\citenamefont{Epelbaum,
  Gl\"ockle, and Mei\ss{}ner}}]{EpGl05}
\bibinfo{author}{\bibfnamefont{E.}~\bibnamefont{Epelbaum}},
  \bibinfo{author}{\bibfnamefont{W.}~\bibnamefont{Gl\"ockle}},
  \bibnamefont{and} \bibinfo{author}{\bibfnamefont{U.-G.}
  \bibnamefont{Mei\ss{}ner}}, \bibinfo{journal}{Nucl. Phys. A}
  \textbf{\bibinfo{volume}{747}}, \bibinfo{pages}{362} (\bibinfo{year}{2005}).

\bibitem[{\citenamefont{Navratil}(2007)}]{Navr07}
\bibinfo{author}{\bibfnamefont{P.}~\bibnamefont{Navr\'atil}},
  \bibinfo{journal}{Few Body Syst.} \textbf{\bibinfo{volume}{41}},
  \bibinfo{pages}{117} (\bibinfo{year}{2007}).

\bibitem[{\citenamefont{Navr\'atil et~al.}(2007)\citenamefont{Navr\'atil,
  Gueorguiev, Vary, Ormand, and Nogga}}]{NaGu07}
\bibinfo{author}{\bibfnamefont{P.}~\bibnamefont{Navr\'atil}},
  \bibinfo{author}{\bibfnamefont{V.~G.} \bibnamefont{Gueorguiev}},
  \bibinfo{author}{\bibfnamefont{J.~P.} \bibnamefont{Vary}},
  \bibinfo{author}{\bibfnamefont{W.~E.} \bibnamefont{Ormand}},
  \bibnamefont{and} \bibinfo{author}{\bibfnamefont{A.}~\bibnamefont{Nogga}},
  \bibinfo{journal}{Phys. Rev. Lett.} \textbf{\bibinfo{volume}{99}},
  \bibinfo{pages}{042501} (\bibinfo{year}{2007}).

\bibitem[{\citenamefont{Maris et~al.}(2011)\citenamefont{Maris, Vary, Navratil,
  Ormand, Nam, and Dean}}]{MaVa11}
\bibinfo{author}{\bibfnamefont{P.}~\bibnamefont{Maris}},
  \bibinfo{author}{\bibfnamefont{J.~P.} \bibnamefont{Vary}},
  \bibinfo{author}{\bibfnamefont{P.}~\bibnamefont{Navr\'atil}},
  \bibinfo{author}{\bibfnamefont{W.~E.} \bibnamefont{Ormand}},
  \bibinfo{author}{\bibfnamefont{H.}~\bibnamefont{Nam}}, \bibnamefont{and}
  \bibinfo{author}{\bibfnamefont{D.~J.} \bibnamefont{Dean}},
  \bibinfo{journal}{Phys. Rev. Lett.} \textbf{\bibinfo{volume}{106}},
  \bibinfo{pages}{202502} (\bibinfo{year}{2011}).

\bibitem[{\citenamefont{Maris et~al.}(2012)\citenamefont{Maris, Vary, and
  Navratil}}]{MaVa12}
\bibinfo{author}{\bibfnamefont{P.}~\bibnamefont{Maris}},
  \bibinfo{author}{\bibfnamefont{J.~P.} \bibnamefont{Vary}}, \bibnamefont{and}
  \bibinfo{author}{\bibfnamefont{P.}~\bibnamefont{Navr\'atil}},
  \bibinfo{journal}{arXiv: 1205.5686}  (\bibinfo{year}{2012}).

\bibitem[{\citenamefont{Roth et~al.}(2011)\citenamefont{Roth, Langhammer,
  Calci, Binder, and Navr{\'a}til}}]{RoLa11}
\bibinfo{author}{\bibfnamefont{R.}~\bibnamefont{Roth}},
  \bibinfo{author}{\bibfnamefont{J.}~\bibnamefont{Langhammer}},
  \bibinfo{author}{\bibfnamefont{A.}~\bibnamefont{Calci}},
  \bibinfo{author}{\bibfnamefont{S.}~\bibnamefont{Binder}}, \bibnamefont{and}
  \bibinfo{author}{\bibfnamefont{P.}~\bibnamefont{Navr{\'a}til}},
  \bibinfo{journal}{Phys. Rev. Lett.} \textbf{\bibinfo{volume}{107}},
  \bibinfo{pages}{072501} (\bibinfo{year}{2011}).

\bibitem[{\citenamefont{Bernard et~al.}(2008)\citenamefont{Bernard, Epelbaum,
  Krebs, and Mei\ss{}ner}}]{BeEp08}
\bibinfo{author}{\bibfnamefont{V.}~\bibnamefont{Bernard}},
  \bibinfo{author}{\bibfnamefont{E.}~\bibnamefont{Epelbaum}},
  \bibinfo{author}{\bibfnamefont{H.}~\bibnamefont{Krebs}}, \bibnamefont{and}
  \bibinfo{author}{\bibfnamefont{U.-G.} \bibnamefont{Mei\ss{}ner}},
  \bibinfo{journal}{Phys. Rev. C} \textbf{\bibinfo{volume}{77}},
  \bibinfo{pages}{064004} (\bibinfo{year}{2008}).

\bibitem[{\citenamefont{Bernard et~al.}(2011)\citenamefont{Bernard, Epelbaum,
  Krebs, and Meissner}}]{BeEp11}
\bibinfo{author}{\bibfnamefont{V.}~\bibnamefont{Bernard}},
  \bibinfo{author}{\bibfnamefont{E.}~\bibnamefont{Epelbaum}},
  \bibinfo{author}{\bibfnamefont{H.}~\bibnamefont{Krebs}}, \bibnamefont{and}
  \bibinfo{author}{\bibfnamefont{U.-G.} \bibnamefont{Mei\ss{}ner}},
  \bibinfo{journal}{Phys. Rev. C} \textbf{\bibinfo{volume}{84}},
  \bibinfo{pages}{054001} (\bibinfo{year}{2011}).
  
\bibitem[{\citenamefont{Skibinski et~al.}(2011)\citenamefont{Skibinski, Golak,
  Topolnicki, Witala, Epelbaum, Gloeckle, Krebs, Nogga, and Kamada}}]{SkGo11}
\bibinfo{author}{\bibfnamefont{R.}~\bibnamefont{Skibinski}},
  \bibinfo{author}{\bibfnamefont{J.}~\bibnamefont{Golak}},
  \bibinfo{author}{\bibfnamefont{K.}~\bibnamefont{Topolnicki}},
  \bibinfo{author}{\bibfnamefont{H.}~\bibnamefont{Witala}},
  \bibinfo{author}{\bibfnamefont{E.}~\bibnamefont{Epelbaum}},
  \bibinfo{author}{\bibfnamefont{W.}~\bibnamefont{Gloeckle}},
  \bibinfo{author}{\bibfnamefont{H.}~\bibnamefont{Krebs}},
  \bibinfo{author}{\bibfnamefont{A.}~\bibnamefont{Nogga}}, \bibnamefont{and}
  \bibinfo{author}{\bibfnamefont{H.}~\bibnamefont{Kamada}},
  \bibinfo{journal}{Phys. Rev. C} \textbf{\bibinfo{volume}{84}},
  \bibinfo{pages}{054005} (\bibinfo{year}{2011}).

\bibitem[{\citenamefont{Krebs et~al.}(2007)\citenamefont{Krebs, Epelbaum, and
  Mei{\ss}ner}}]{KrEp07}
\bibinfo{author}{\bibfnamefont{H.}~\bibnamefont{Krebs}},
  \bibinfo{author}{\bibfnamefont{E.}~\bibnamefont{Epelbaum}}, \bibnamefont{and}
  \bibinfo{author}{\bibfnamefont{U.-G.} \bibnamefont{Mei{\ss}ner}},
  \bibinfo{journal}{Eur. Phys. J. A} \textbf{\bibinfo{volume}{32}},
  \bibinfo{pages}{127} (\bibinfo{year}{2007}).

\bibitem[{\citenamefont{Hagen et~al.}(2012{\natexlab{b}})\citenamefont{Hagen,
  Hjorth-Jensen, Jansen, Machleidt, and Papenbrock}}]{HaHj12a}
\bibinfo{author}{\bibfnamefont{G.}~\bibnamefont{Hagen}},
  \bibinfo{author}{\bibfnamefont{M.}~\bibnamefont{Hjorth-Jensen}},
  \bibinfo{author}{\bibfnamefont{G.~R.} \bibnamefont{Jansen}},
  \bibinfo{author}{\bibfnamefont{R.}~\bibnamefont{Machleidt}},
  \bibnamefont{and}
  \bibinfo{author}{\bibfnamefont{T.}~\bibnamefont{Papenbrock}},
  \bibinfo{journal}{Phys. Rev. Lett.} \textbf{\bibinfo{volume}{108}},
  \bibinfo{pages}{242501} (\bibinfo{year}{2012}{\natexlab{b}}).

\bibitem[{\citenamefont{Roth et~al.}(2012)\citenamefont{Roth, Binder, Vobig,
  Calci, Langhammer, and Navr\'atil}}]{RoBi12}
\bibinfo{author}{\bibfnamefont{R.}~\bibnamefont{Roth}},
  \bibinfo{author}{\bibfnamefont{S.}~\bibnamefont{Binder}},
  \bibinfo{author}{\bibfnamefont{K.}~\bibnamefont{Vobig}},
  \bibinfo{author}{\bibfnamefont{A.}~\bibnamefont{Calci}},
  \bibinfo{author}{\bibfnamefont{J.}~\bibnamefont{Langhammer}},
  \bibnamefont{and}
  \bibinfo{author}{\bibfnamefont{P.}~\bibnamefont{Navr\'atil}},
  \bibinfo{journal}{Phys. Rev. Lett.} \textbf{\bibinfo{volume}{109}},
  \bibinfo{pages}{052501} (\bibinfo{year}{2012}).

\bibitem[{\citenamefont{Jurgenson et~al.}(2009)\citenamefont{Jurgenson,
  Navr\'{a}til, and Furnstahl}}]{JuNa09}
\bibinfo{author}{\bibfnamefont{E.~D.} \bibnamefont{Jurgenson}},
  \bibinfo{author}{\bibfnamefont{P.}~\bibnamefont{Navr\'{a}til}},
  \bibnamefont{and} \bibinfo{author}{\bibfnamefont{R.~J.}
  \bibnamefont{Furnstahl}}, \bibinfo{journal}{Phys. Rev. Lett.}
  \textbf{\bibinfo{volume}{103}}, \bibinfo{pages}{082501}
  (\bibinfo{year}{2009}).

\bibitem[{\citenamefont{Hagen et~al.}(2007)\citenamefont{Hagen, Papenbrock,
  Dean, Schwenk, Nogga, W{\l}och, and Piecuch}}]{HaPa07}
\bibinfo{author}{\bibfnamefont{G.}~\bibnamefont{Hagen}},
  \bibinfo{author}{\bibfnamefont{T.}~\bibnamefont{Papenbrock}},
  \bibinfo{author}{\bibfnamefont{D.~J.} \bibnamefont{Dean}},
  \bibinfo{author}{\bibfnamefont{A.}~\bibnamefont{Schwenk}},
  \bibinfo{author}{\bibfnamefont{A.}~\bibnamefont{Nogga}},
  \bibinfo{author}{\bibfnamefont{M.}~\bibnamefont{W{\l}och}}, \bibnamefont{and}
  \bibinfo{author}{\bibfnamefont{P.}~\bibnamefont{Piecuch}},
  \bibinfo{journal}{Phys. Rev. C} \textbf{\bibinfo{volume}{76}},
  \bibinfo{pages}{034302} (\bibinfo{year}{2007}).

\bibitem[{\citenamefont{Shavitt and Bartlett}(2009)}]{ShBa09}
\bibinfo{author}{\bibfnamefont{I.}~\bibnamefont{Shavitt}} \bibnamefont{and}
  \bibinfo{author}{\bibfnamefont{R.}~\bibnamefont{Bartlett}},
  \emph{\bibinfo{title}{Many-Body Methods in Chemistry and Physics}}
  (\bibinfo{publisher}{Cambridge University Press}, \bibinfo{year}{2009}).

\bibitem[{\citenamefont{Hagen et~al.}(2010)\citenamefont{Hagen, Papenbrock,
  Dean, and Hjorth-Jensen}}]{HaPa10}
\bibinfo{author}{\bibfnamefont{G.}~\bibnamefont{Hagen}},
  \bibinfo{author}{\bibfnamefont{T.}~\bibnamefont{Papenbrock}},
  \bibinfo{author}{\bibfnamefont{D.~J.} \bibnamefont{Dean}}, \bibnamefont{and}
  \bibinfo{author}{\bibfnamefont{M.}~\bibnamefont{Hjorth-Jensen}},
  \bibinfo{journal}{Phys. Rev.} \textbf{\bibinfo{volume}{C82}},
  \bibinfo{pages}{034330} (\bibinfo{year}{2010}).

\bibitem[{\citenamefont{Roth et~al.}()\citenamefont{Roth, Calci, Langhammer,
  and Binder}}]{RoCa13}
\bibinfo{author}{\bibfnamefont{R.}~\bibnamefont{Roth}},
  \bibinfo{author}{\bibfnamefont{A.}~\bibnamefont{Calci}},
  \bibinfo{author}{\bibfnamefont{J.}~\bibnamefont{Langhammer}},
  \bibnamefont{and} \bibinfo{author}{\bibfnamefont{S.}~\bibnamefont{Binder}},
  \bibinfo{note}{in preparation}.

\bibitem[{\citenamefont{Taube and Bartlett}(2008{\natexlab{a}})}]{TaRo08}
\bibinfo{author}{\bibfnamefont{A.~G.} \bibnamefont{Taube}} \bibnamefont{and}
  \bibinfo{author}{\bibfnamefont{R.~J.} \bibnamefont{Bartlett}},
  \bibinfo{journal}{The Journal of Chemical Physics}
  \textbf{\bibinfo{volume}{128}}, \bibinfo{eid}{044110} (\bibinfo{year}{2008}{\natexlab{a}}).

\bibitem[{\citenamefont{Taube and Bartlett}(2008{\natexlab{b}})}]{TaRo08-2}
\bibinfo{author}{\bibfnamefont{A.~G.} \bibnamefont{Taube}} \bibnamefont{and}
  \bibinfo{author}{\bibfnamefont{R.~J.} \bibnamefont{Bartlett}},
  \bibinfo{journal}{The Journal of Chemical Physics}
  \textbf{\bibinfo{volume}{128}}, \bibinfo{eid}{044111} (\bibinfo{year}{2008}{\natexlab{b}}).

\end{thebibliography}

\end{document}